\documentclass[nofootinbib,twocolumn]{revtex4}
\usepackage{graphicx}
\usepackage{amssymb}
\usepackage{amsmath}
\usepackage{bm}
\usepackage{hyperref}
\begin{document}

\title{\boldmath {\it Gauging} the cosmic acceleration with recent type Ia supernovae data sets}

\author{Hermano Velten}%
 \email{velten@pq.cnpq.br}
\affiliation{N\'ucleo Cosmo-ufes \& Departamento de F\'isica, CCE, Universidade Federal do Esp\'irito Santo, 29075-910, Vit\'oria-ES, Brazil}
\author{Syrios Gomes}%
 \email{syriosgs@gmail.com}
\affiliation{N\'ucleo Cosmo-ufes \& Departamento de F\'isica, CCE, Universidade Federal do Esp\'irito Santo, 29075-910, Vit\'oria-ES, Brazil}

\author{Vinicius C. Busti}
 \email{busti@sas.upenn.edu}
\affiliation{
 Department of Physics and Astronomy, University of Pennsylvania, Philadelphia, PA 19104, USA}
\affiliation{Departamento de F\'isica Matem\'atica, Instituto de F\'isica, Universidade de S\~ao Paulo, 05508-090, S\~ao Paulo - SP, Brasil}


\begin{abstract}
\noindent
We revisit a model-independent estimator for cosmic acceleration based on type Ia supernovae distance measurements. This approach does not rely on any specific theory for gravity, energy content or parameterization for the scale factor or deceleration parameter and is based on falsifying the {\it null hypothesis} that the Universe never expanded in an accelerated way. By generating mock catalogues of known cosmologies we test the robustness of this estimator establishing its limits of applicability. We detail the pros and cons of such approach. For example, we find that there are specific counterexamples in which the estimator wrongly provides evidence against acceleration in accelerating cosmologies. The dependence of the estimator on the $H_0$ value is also discussed. Finally, we update the evidence for acceleration using the recent UNION2.1 and JLA samples. Contrary to recent claims, available data strongly favors an accelerated expansion of the Universe in complete agreement with the standard $\Lambda$CDM model.
\end{abstract}

\maketitle

\section{Introduction}

Distance measurements to type Ia supernovae (SNe Ia) at high redshifts led to the astonishing discovery of the cosmic acceleration in 1998 \cite{Perlmutter:1998np,Riess:1998cb}. Within a general relativistic based description of gravity, with the background expansion equipped with an FLRW (Friedmann-Lemaitre-Robertson-Walker) metric it is necessary the inclusion of some unknown form of energy density (dark energy) responsible for driving such dynamics. The simplest explanation relies on the Einstein's cosmological constant $\Lambda$, which is equivalent at the background level to a fluid with an equation of state parameter $w_{de}=p_{de}/\rho_{de}=-1$. However, there are also alternative descriptions for the accelerated expansion of the universe as for example modifications of Einstein's gravity, backreaction mechanisms or viscous effects among many others. In all the above approaches (including the standard one) the evidence for acceleration appears from fitting data and realizing that the parameter space of such models leading to the accelerated expansion is the statistically favored one. 

Another way to probe the acceleration is to perform kinematical tests with data where no assumptions about the gravitational sector or the material content of the universe are made. Within this class of tests one can cite parameterizations of the deceleration parameter $q(z)$ \cite{Elgaroy:2006tp,
Shafieloo:2009ti}, the scale factor $a(t)$ \cite{Wang:2005yaa} or the expansion rate $H(z)$ \cite{John:2005bz,Nair:2012bs}, as well as cosmographical tests employing a series expansion in the redshift $z$ \cite{review_cosmography,Luongo:2015zgq,cosm16}, however see \cite{busti2015} for limitations of such an approach. 

In this work we revisit another model-independent estimator for accelerated expansion described in great detail by Schwarz and Seikel \cite{Seikel:2007pk,Seikel:2008ms}. The idea is based on falsifying the {\it null hypothesis} that the universe never experienced an accelerated expansion. Although such estimator does not provide the moment at which the transition to acceleration takes place (such criticism has been discussed in Ref. \cite{Mortsell:2008yu}) the analyses with the GOLD 
\cite{Riess:2006fw}, ESSENCE\cite{WoodVasey:2007jb} and UNION \cite{Kowalski:2008ez} SNe Ia data sets have provided strong evidence in favor of acceleration \cite{Seikel:2007pk,Seikel:2008ms}. 

Our aim in this work is twofold: First, we check the robustness of the estimator by testing it against mock catalogues of known cosmological expansions (e.g., $\Lambda$CDM, Einstein-de Sitter, Milne model and 3 other models which are not accelerated today but were accelerated in the past). This analysis allows us to understand the estimator and to {\it gauge} the level of accuracy expected for the statistical evidence (given in $\sigma$ levels) obtained with actual data sets. Second, we update the results of such estimator for recent SNe Ia data sets like the UNION2.1 \cite{Suzuki:2011hu} and the Joint Light-Curve analysis (JLA) \cite{Betoule:2014frx} samples. 

Concerning the confrontation of a model independent estimator for acceleration with the recent JLA sample, it is worth noting that
recently Ref. \cite{Dam:2017xqs} has argued that the timescape model (with insignificant acceleration rate) fits the JLA sample with a likelihood that is
statistically indistinguishable from the standard cosmological model. Even more intriguing, Ref.\cite{Nielsen:2015pga} has pointed out that the JLA sample provides only marginal evidence for acceleration. However, Ref. \cite{Rubin:2016iqe} has criticized the former result by arguing that the statistical model used in Ref.\cite{Nielsen:2015pga} is deficient to account for
changes in the observed SN light-curve parameter distributions with redshift. According to Ref. \cite{Rubin:2016iqe} evidence for acceleration using SNe Ia only is $11.2 \sigma$ in a flat universe. Thus, in our work we are also able to revisit this discussion by analyzing the evidence for acceleration in the JLA sample from a different perspective.

In the next section we review the estimator using it in section 3 with the UNION2.1 \cite{Suzuki:2011hu} and in section 4 with the Joint-Lightcurve-Analysis (JLA) \cite{Betoule:2014frx} samples. We conclude in the final section.

\section{A model-independent estimator for cosmic acceleration}

We review in this section the estimator developed in Ref. \cite{Seikel:2007pk} by Seikel \& Schwarz. The main idea here is to provide a quantitative measure of the accelerated dynamics of the universe in a model-independent way. The construction of such estimator is based on the definition of the deceleration parameter
\begin{equation}\label{qz}
q(z) = \frac{H^{\prime}(z)}{H(z)} (1+z) - 1 
\end{equation}
where the prime denotes derivative with respect to the redshift. An accelerated background expansion at a certain redshift is indicated if $q(z)<0$. 

In the the standard cosmology one expects that the dynamical evolution of the universe underwent a transition from the decelerated phase to the accelerated one at some transition redshift $z_t$. Deep in the matter dominated epoch the deceleration parameter assumes the value $\sim 0.5$ (similarly to the Einstein-de Sitter universe). As the effect of dark energy (or modified gravity) becomes relevant for the expansion in comparison to the matter density, then $q(z)$ turns to negative values. For the standard cosmology $q(z)$ is seen in the black line of Fig. {\ref{Figq}}.

From Eq. (\ref{qz}) one can obtain the expansion rate $H(z)$ as a function of the deceleration parameter according to the integral equation
\begin{equation}
{\rm ln}\,\frac{H(z)}{H_0}=\int^z_0 \frac{1+q(\tilde{z})}{1+\tilde{z}} d\tilde{z},
\end{equation}
where $H_0$ is the Hubble constant.

The {\it null hypothesis} proposed in \cite{Seikel:2007pk} is that the universe has never expanded in an accelerated way i.e., $q(z)>0 \,\forall \,z$. Hence, the direct consequence of applying this inequality to the above equation is

\begin{equation}
{\rm ln}\,\frac{H(z)}{H_0}\geq\int^z_0\frac{1}{1+\tilde{z}}d\tilde{z}={\rm ln}\,(1+z),
\end{equation}
which is equivalent to 
\begin{equation}\label{hzcondition}
H(z)\geq H_0 (1+z).
\end{equation}
Therefore, from the above result one can infer whether or not the universe experienced any event of accelerated expansion from direct measurements of the Hubble rate. This can be achieved for example using the so called cosmic chronometers, which are galaxies supposed to passively evolve in a certain redshift range $\Delta z$. Then, estimation of the stellar ages ($\Delta t$) in such objects lead to an estimation of $H(z)= -1/(1+z) \, dz/dt \sim -1/(1+z) \, \Delta z / \Delta t$. However the available number of $H(z)$ data is limited to a few dozens and the quality (in terms of the associated errors) is low.

In order to assess information on the background expansion of the universe  the SNe Ia data is the most reliable observational tool. The quantity and quality of SNe Ia data have substantially increased in the last years and ongoing surveys will drastically improve SNe Ia samples in the near future. The crucial definition in supernovae cosmology is the luminosity distance. In a flat FLRW universe it reads
\begin{equation}
d_L(z)= (1+z) \int^{z}_0 \frac{d\tilde{z}}{H(\tilde{z})}.
\end{equation}
Now, in order to apply the inequality (\ref{hzcondition}) in the context of supernovae data the definition $d_L$ turns into 
\begin{equation}
d_L\leq (1+z) \frac{1}{H_0} \int^z_0 \frac{d\tilde{z}}{1+\tilde{z}}=(1+z)\frac{1}{H_0} \ln (1+z). 
\end{equation}

The luminosity distance calculated in some theoretical model under investigation is compared to the observed quantities via the definition of the observed distance modulus
\begin{equation}
\mu = m-M= 5 \log (d_L / {\rm Mpc}) + 25,
\end{equation}
where $M$ and $m$ are the absolute and apparent magnitudes, respectively.

For each supernova $i$ in the sample we can define the quantity
\begin{eqnarray}
\Delta \mu_{obs} (z_i) &=& \mu_{obs}(z_i) - \mu(q=0) \\ \nonumber
&=& \mu_{obs} (z_i)-5{\rm log}\left[\frac{1}{H_0}(1+z_i)\ln (1+z_i)\right]-25,
\label{eqDeltaobs}
\end{eqnarray}
which is the difference between its observed distance modulus $\mu_{obs}(z_i)$ and the distance modulus of a universe with constant deceleration parameter $q=0$ at the redshift $z_i$.


The null hypothesis behind the estimator of Ref. \cite{Seikel:2007pk} is that the universe never expanded in an accelerated way which corresponds to $\Delta \mu_{obs} {\leq} 0$ for each supernova. Oppositely, positive $\Delta \mu_{\rm obs}$ values indicate acceleration. 

The face value of $\Delta \mu_{obs}$ is of limited interest if its associated error $\sigma_i$ is not included. For each sample used in this work (UNION2.1 and JLA) we will obtain the error in the distance moduli of each supernovae $i$ ($\sigma_i$) from the available covariance matrix $C$ of the data. 

One way to apply the estimator is via the the so called ``single SNe Ia analysis" which corresponds to compute the quantity $\Delta \mu_{obs}$ for each SN individually. Although the single SN analysis presents some interesting results, it is however of limited statistical interest. A more reliable analysis of the estimator $\Delta \mu_{obs}$ is obtained with the so called ``averaged SNe Ia analysis". In this analysis we group a number N of SNe defining the mean value

\begin{equation}
\overline{\Delta\mu}=\frac{\sum_{i=1}^{N} g_i \,\Delta\mu_{obs}(z_i)}{\sum_{i=1}^{N} g_i},
\end{equation}
where the factor $g_i = 1/ \sigma^{2}_i$ enables data points with smaller errors contribute more to the average. 

The standard deviation of the mean value is defined by
\begin{equation}
\sigma_{\overline{\Delta\mu}}=\left[\frac{\sum^N_{i=1} g_i \left[\Delta \mu_{obs} \left(z_i\right) - \overline{\Delta\mu}\right]^2}{(N-1)\sum^N_{i=1}g_i}\right]^{1/2}.
\label{sigmadeltamu}
\end{equation}

For example, using (\ref{sigmadeltamu}) , Ref. \cite{Seikel:2008ms} points out averaged statistical evidences for acceleration such that $4.3\sigma$ for the GOLD sample and $7.2\sigma$ using the 2008 UNION sample, both assuming a flat expansion.

\section{The UNION2.1 data set}

\subsection{Understanding the estimator}

As a preliminary study we investigate in more detail the robustness and reliability of the estimator. Actually, we want to verify the outcomes concerning the averaged analysis desiring a better understanding about obtained for the evidence $\overline{\Delta\mu} / \sigma_{\overline{\Delta\mu}}$. In some sense in this subsection we calibrate the ``averaged SNe Ia analysis".



Let us simulate catalogs for given cosmologies in which we know in advance the state of acceleration for every redshift. Then we confront the simulated data with the predictions of the estimator for a given actual catalog. We shall use the redshift distribution of UNION2.1 sample as our reference. 

The models we adopt here are based on the following flat FLRW expansion 
\begin{equation}
\frac{H^2(z)}{H_0^2}= \Omega_{m0} (1+z)^3 + (1-\Omega_{m0})e^{3\int^z_0 \frac{1+w(z^{\prime})}{1+z^{\prime}}dz^{\prime}}.
\end{equation}
The relevant models are the $\Lambda$CDM (with $w=-1$ and $\Omega_{m0}=0.3$), a pure matter dominated Einstein-de Sitter model ($\Omega_{m0}=1$) and the Milne's model.

\begin{figure}
\centering
\includegraphics[width=0.45\textwidth]{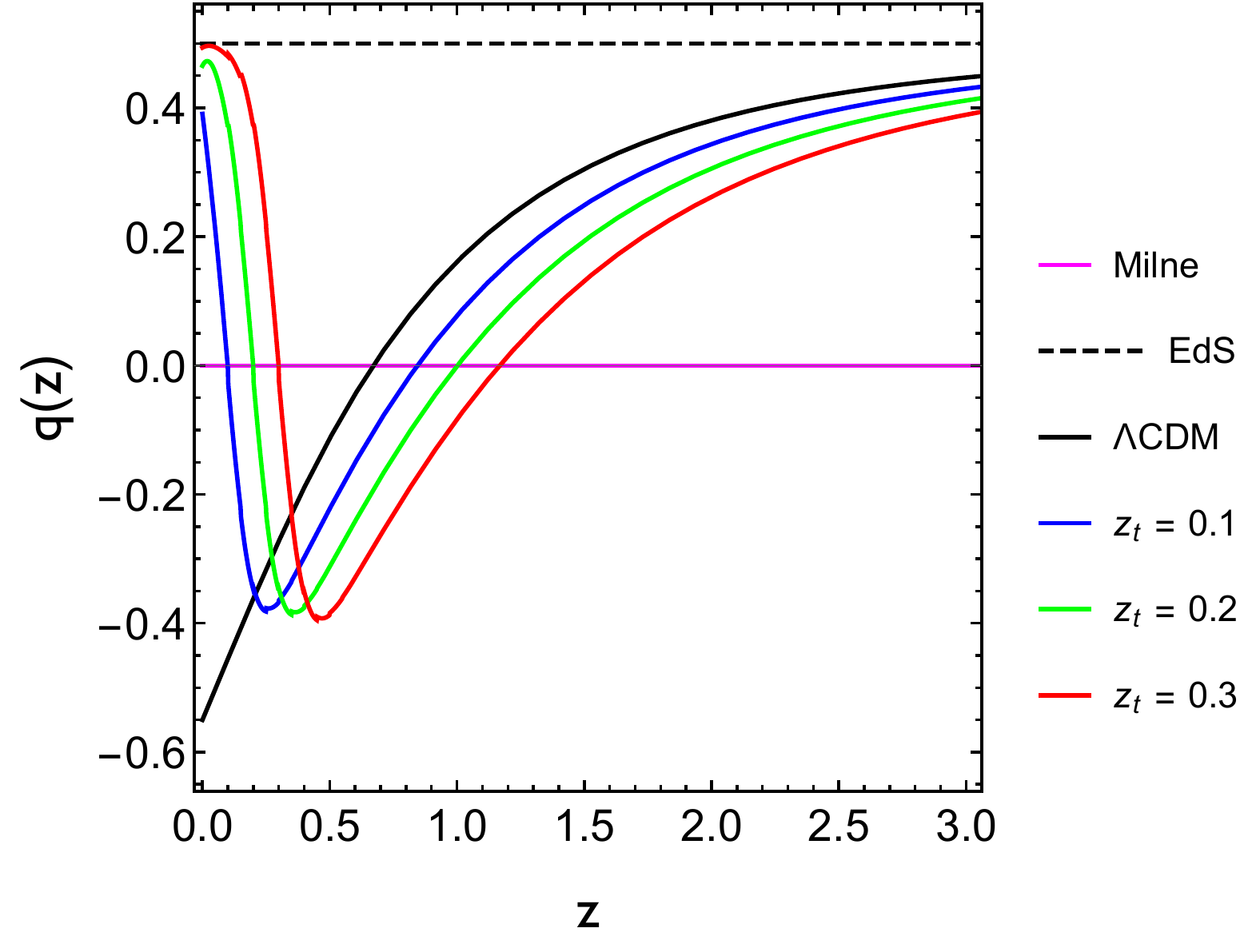}
\caption{Evolution of the deceleration parameter $q(z)$ as a function of the redshift for different models adopted in this work.}
\label{Figq}
\end{figure}

As shown in Fig. \ref{Figq}, in terms of the deceleration parameter, the $\Lambda$CDM model promotes a smooth transition from a decelerated universe with $q (z>>1) = +0.5$ to a recent accelerated expansion $q(z=0) = -0.68$. The transition redshift at which $q (z_{ac})=0$ occurs at $z_{ac} = 0.67$. For the EdS model the universe is always decelerating at a constant rate, i.e., $q=+0.5 \,\forall \, z$. The Milne's model corresponds to a constant expansion rate given by $q=0 \,\forall \, z$.

\begin{figure}
\centering
\includegraphics[width=0.45\textwidth]{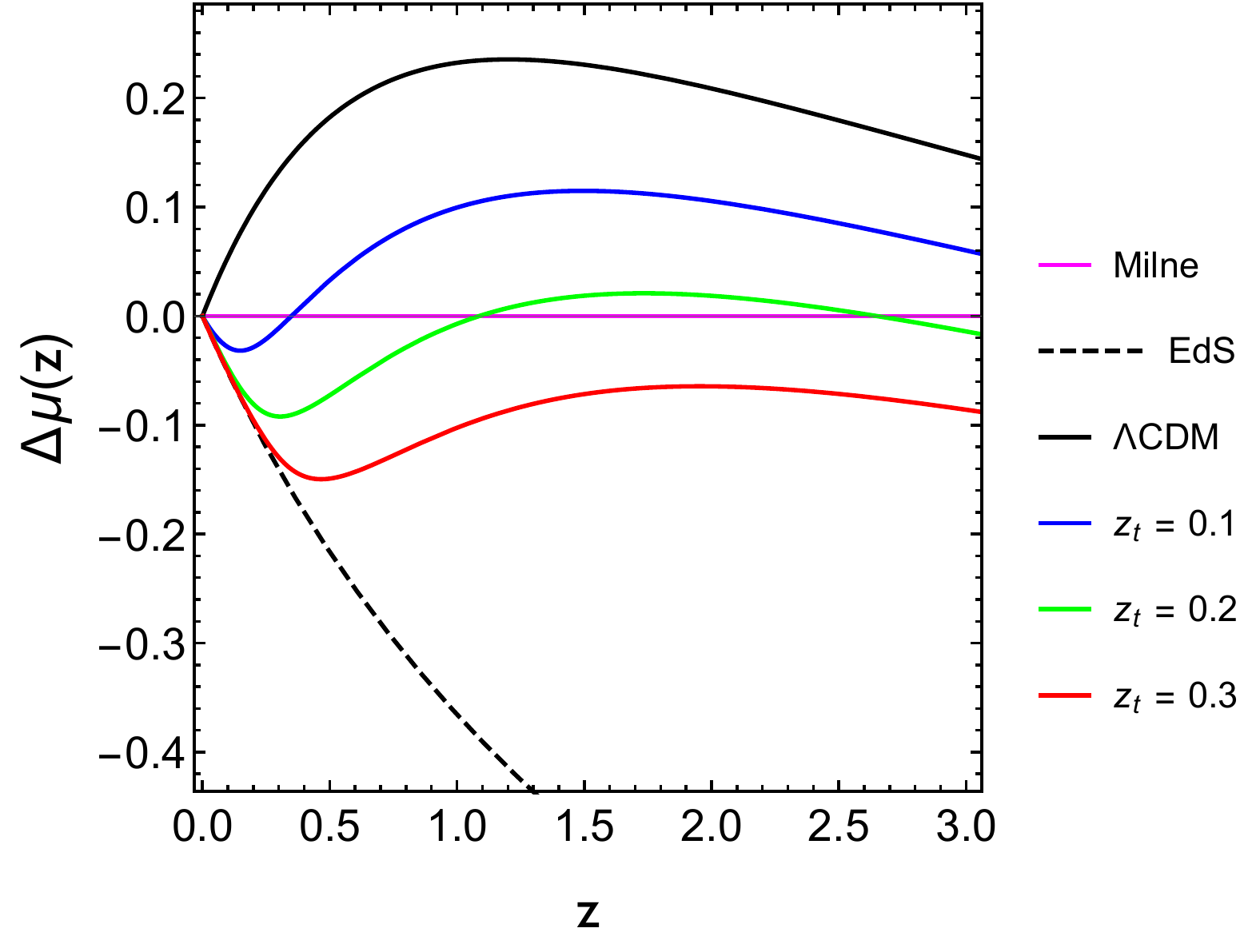}
\caption{Evolution of $\Delta \mu$ as a function of the redshift for different models adopted in this work.}
\label{FigDeltamuz}
\end{figure}

In addition, in order to check the ability of the estimator with non-usual expansions it is also interesting to study cosmologies in which, after transiting from the decelerated EdS phase to the accelerated one, there is a transition back to a decelerated phase as for example models based on the ansatz \cite{Shafieloo:2009ti}
\begin{equation}
w(z)= -\frac{1+{\rm Tanh}[(z-z_t)\Delta]}{2}.
\label{wzt}
\end{equation}
In the above expression $z_t$ denotes the redshift at which the expansion turns to be decelerated and $\Delta$ the duration of the accelerated epoch. 
We adopt three other cases where $z_t=0.1, 0.2$ and $0.3$ all assuming $\Delta=10$ (see Ref. \cite{Shafieloo:2009ti} for details). All such latter models had indeed a phase of accelerated expansion in the past but the current (at $z=0$) expansion is decelerated.
Therefore, we will work with six different cosmological models. Apart from the deceleration parameter we also show in Fig.\ref{FigDeltamuz} the expected value for $\Delta \mu$ for each model. 

We proceed our analysis by asking what should be the evidence value $\overline{\Delta \mu}$ / $\sigma_{\overline{\Delta\mu}}$ of the estimator in each of such cosmologies. In some sense, we try to quantify the information contained in Fig. \ref{FigDeltamuz}.   

At this point it is necessary to point out a cautionary remark. The deceleration parameter $q(z)$ does not depend on the today's value of the Hubble expansion $H_0$. Note however that the estimator $\Delta \mu$ does depend on $H_0$. This fact is related to the current discussion on the tension about the $H_0$ value. The $H_0$ value inferred with cosmological data by fitting the standard $\Lambda$CDM cosmology leads to a lower value $H_0=67.31 \pm 0.96$ km s$^{-1}$ Mpc$^{-1}$ \cite{ADEPlanck} than the one directly obtained from local measurements $H_0=73.24 \pm 1.74$ km s$^{-1}$ Mpc$^{-1}$ \cite{RiessH0}. Then, in order to use the estimator as model independent as possible we use the latter value.

Table \ref{taUNION} shows our simulated results of the estimator for known cosmologies. We have generated $1000$ different realizations of UNION2.1-like Hubble diagrams i.e., each catalog has 580 data points with the same redshift distribution as the UNION2.1 sample. Given a cosmological model the distance modulus $\mu$ is generated obeying a Gaussian distribution around the exact theoretical value. Each generated $\mu_i$ at a redshift $z_i$ has the same error as the observed one of the UNION2.1 sample at that redshift. The result $\overline{\Delta \mu}$ / $\sigma_{\overline{\Delta\mu}} = 12.77 \pm 0.95$ shown for the $\Lambda$CDM model corresponds to the mean evidence ($12.77$) over all realizations with the corresponding distribution standard deviation ($0.95$). Later on, this result should be compared directly to the one corresponding to actual data. We will perform this analysis in the next subsection (see Table \ref{TableAveUNION}). 

Still in Table \ref{taUNION}, the result for the Milne's model $\overline{\Delta \mu}$ / $\sigma_{\overline{\Delta\mu}} = 0.05 \pm 0.95$ provides a good indication for the robustness of the estimator. It is worth remembering that the exact value for this case is $\overline{\Delta \mu}$ / $\sigma_{\overline{\Delta\mu}} = 0 $. Also, Table \ref{taUNION} indicates that the typical standard deviation value is around the unity for all models studied.
Interesting to notice is an apparent failure of the estimator for the cases $z_t=0.2$ and $z_t=0.3$. The negative values for $\overline{\Delta \mu}$ / $\sigma_{\overline{\Delta\mu}}$ (taking into account the variance) clearly do not indicate the existence of the accelerated phase and are examples of situations in which the estimator fails to provide the correct answer.

Since we have found with the models $z_t=0.2$ and $z_t=0.3$ examples in which the use of estimator to the total sample fails we investigate in deeper detail a binned sample analysis. We show in Table \ref{taBinMOck} the evidence $\overline{\Delta \mu}$ / $\sigma_{\overline{\Delta\mu}}$ per bin width $\Delta z = 0.2$ for the $z_t$-models only. The second column presents the number of supernovae in each bin. The evidence presented in the remaining columns is the averaged evidence over the $N=1000$ mock realizations. 

It is worth noting that the use of high redshift data only ($z>1.2$) in the analysis with the model $z_t=0.2$ favored acceleration (as expected from Fig.\ref{FigDeltamuz}). This reflects the dangers in trusting the estimator if using selectively SNe data.


Now, in order to check how the available redshift range of the observed SNe impacts the final outcome of the estimator we promote a second analysis in which we generate for each mock sample $1500$ SNe equally distributed in the redshift range $0<z<1.5$. With this example there will be no proper comparison with the actual UNION2.1 data set but this analysis is useful to understand how the indicator works. Each generated distance modulus has the same constant error $\sigma_i = 0.15$. The averaged evidences for $N=1000$ sample realizations of such redshift distribution are shown in Table \ref{ta1500}. We obtain again the expected result for the Milne's model. The high value for the averaged evidence in the $\Lambda$CDM case $\overline{\Delta \mu}$ / $\sigma_{\overline{\Delta\mu}}=44.03$ occurs because now there are more SNe data around the redshift $z\sim 1.2$ where $\Delta \mu$ is expected to peak at its maximum value (see Fig. \ref{FigDeltamuz}).

\begin{table}
\centering\caption{Averaged evidence for acceleration using N=1000 mock catalogs. Models are plotted in Fig.1. For each catalog there are 580 SNe with the same redshift distribution as the UNION2.1 sample.}
{\begin{tabular} {c||c||c}
Model & $\overline{\Delta \mu}$ / $\sigma_{\overline{\Delta\mu}}$   & Std. Dev. \\
\hline \hline 
$\Lambda$CDM &  12.77 & $\pm$ 0.94  \\
Einstein-de Sitter&  -13.95 & $\pm$ 0.97\\
Milne (q=0 $\forall$ z) &  0.05 & $\pm$ 0.95  \\
$z_t$=0.1 &  -0.15& $\pm$ 0.97  \\
$z_t$=0.2 &  -7.19& $\pm$ 0.92  \\
$z_t$=0.3 &  -10.97& $\pm$ 1.05  \\
\hline
\end{tabular}} 
\label{taUNION}
\end{table}

\begin{table}
\centering
\caption{Averaged analysis per bin for the mock catalogs generated by the three cosmologies with the transition back to deceleration at $z_t$. The first column shows the redshift bin. The second column the number of SN in each bin. Third, fourth and fifth columns the evidences for the models with 
 $z_t=0.1$, $z_t=0.2$ and $z_t=0.3$, respectively.}
\label{my-label}
\begin{tabular}{ccccccc}
\hline
&Models: & $z_t=0.1$ &$z_t=0.2$ & $z_t=0.3$ \\ \hline
Bin     & \#SN       &   $\overline{\Delta \mu}$ / $\sigma_{\overline{\Delta\mu}}$    & $\overline{\Delta \mu}$ / $\sigma_{\overline{\Delta\mu}}$  & $\overline{\Delta \mu}$ / $\sigma_{\overline{\Delta\mu}}$     \\ \hline \hline
  0.0-0.2  &     230    &  -2.06 &  -3.67 &    -3.99   \\
  0.2-0.4  &      125     &    -0.94     &  -6.12       &    -8.39    \\
  0.4-0.6   &       101    &      1.51 &    -3.48      &    -7.00      \\
  0.6-0.8   &    51      &   2.16    &    -1.55    &  -4.54        \\
  0.8-1.0   &      44     &       2.27   &      -0.54     &     -2.93    \\
  1.0-1.2   &     16     &      1.88   &    0.02      &   -1.68      \\
  1.2-1.41   &     13      &      1.67   &    0.21       &   -1.18     \\ \hline
\end{tabular}
\label{taBinMOck}
\end{table}

\begin{table}
\centering
\caption{Averaged evidence for N=1000 mock catalogs. For each catalog there are 1500 SN equality distributed in the redsfhit range $0<z<1.5$ possessing the same error $\sigma_i=0.15$. }
{\begin{tabular} {c||c||c}
Model & $\overline{\Delta \mu}$ / $\sigma_{\overline{\Delta\mu}}$   & Std. Dev. \\
\hline \hline 
$\Lambda$CDM & 44.00 & $\pm$ 1.63 \\
Einstein-de Sitter& -53.45 & $\pm$1.33\\
Milne (q=0 $\forall$ z) & 0.03 & $\pm$ 1.03  \\
$z_t=0.1$ & 14.05 & $\pm$ 0.99  \\
$z_t=0.2$ & -8.20 & $\pm$ 0.98  \\
$z_t=0.3$ &  -26.31& $\pm$ 1.17  \\
\hline
\end{tabular}}
\label{ta1500}
\end{table}


\subsection{Single and Averaged SN analysis with actual UNION2.1 data}

The observed distance modulus is provided according to
\begin{equation}
\mu_{obs}=m^{\star}_{B}-M_B+\delta \cdot P(m^{true}_{star}< m^{threshold}_{star})+\alpha \cdot X_1 - \beta \cdot \mathcal{C},
\end{equation}
where $m^{\star}_{B}$ is the $B$ band rest-frame observed peak magnitude, $\mathcal{C}$ describes the supernova color at maximum brightness, $X_1$ describes the time stretching of the light-curve and the $M_B$ is the absolute $B-$band magnitude. The parameters $\alpha$, $\beta$, $M_{B}$ and $\delta$ are free and should be inferred via a statistical analysis. This occurs by minimizing the proper $\chi^2$ statistics simultaneously with the free parameters of the cosmological model used in the data fitting.

For the UNION 2.1 sample we adopt the suggested $\mu_{obs}$ x $z$ data with $\alpha=0.121$, $\beta=2.47$, $M_{B}=-19.321$ and $\delta=-0.032$ fitted together with the standard $\Lambda$CDM background cosmology (see Ref. \cite{Suzuki:2011hu}). 

\begin{figure}
\includegraphics[width=0.45\textwidth]{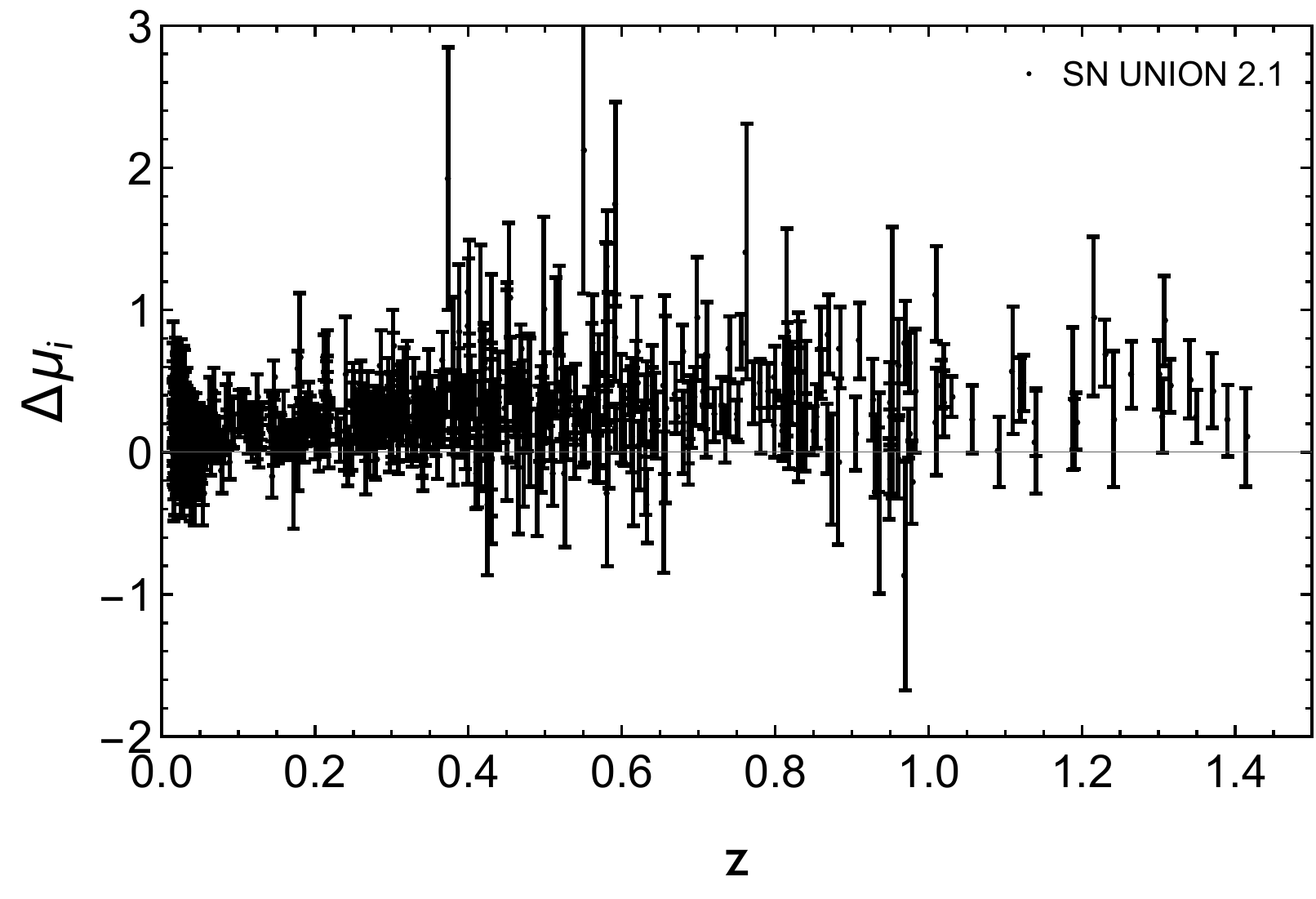}
\caption{Single SNe Ia analysis: $\Delta \mu_i$ for each supernova in the Union 2.1 data set.}
\label{FigSingle}
\end{figure}

According to the single SNe Ia analysis, in Fig. \ref{FigSingle} we show $\Delta \mu_i$ for each supernova in the UNION2.1 sample. Following Ref. \cite{Seikel:2007pk} we adopt a statistical quality control of our sample (control chart). We count SNe in the sample establishing limits for a given control chart i.e., in our case, we want to assert the acceleration at certain statistical confidence level (CL). Values of $\Delta \mu_i$ above an action limit $A_{95} = 1.645 \sigma_i \, \, (A_{99} (99\%) = 2.326 \sigma_i)$ indicates acceleration at $95\% CL \,\,(99\% CL)$.


\begin{table}
\centering\caption{(Single SNe Ia Analysis) Number of Union 2.1 SN indicating acceleration or deceleration. The values in parenthesis are calculated using $H_0=70$ km s$^{-1}$ Mpc$^{-1}$.}
{\begin{tabular} {c||c||c}
Dynamics & Union 2.1 & Union 2.1. \\
 &  & ($z>0.2$) \\
\hline \hline 
Acceleration-$95 \% C.L.$ & 187 (101) & 141 (82)  \\
Acceleration-$99 \% C.L.$ & 87 (34) & 66 (27)  \\
Deceleration-$95 \% C.L.$ & 1 (10) & 0 (1)  \\
Deceleration-$99 \% C.L.$ & 0 (1) & 0 (0)  \\
Total \# of SN & 580 & 350  \\
\hline
\end{tabular}} 
\label{TableSingleUNION}
\end{table}

In Table (\ref{TableSingleUNION}) we show the number of SNe in the UNION2.1 sample presenting acceleration at $95\% CL$ and $99\% CL$. The results considering the total sample ($580$ SNe) are presented in the central column.

Ref. \cite{Seikel:2008ms} also brings the discussion whether or not SNe at low redshifts are trustful for this analysis. It is argued that since most of the nearby SNe have been observed by different projects then different calibrations plagues (by introducing a large systematics) this sub-set. Also, the assumption of homogeneity and isotropy breaks down at scales below a few hundred Mpc then, rather than an evidence for dark energy the Hubble diagram could manifest a violation of the Copernican principle. The results for the sub-sample where all SNe with $z<0.2$ are discarded (it totals now 350 SNe) is shown in the third column of Table (\ref{TableSingleUNION}).

We apply now the averaged evidence for actual data sets. Rather than computing the averaged evidence for the entire sample, one can also present this value for bins of SNe. The grouping criteria can obey either a fixed redshift range $\Delta z$ or a fixed SNe number per bin. The evidence for acceleration in each SN bin is then given by $\overline{\Delta\mu}$ divided by the error $\sigma_{\overline{\Delta\mu}}$.


In table \ref{TableAveUNION} the results are presented considering bins of equal redshift width $\Delta z = 0.2$. The evidence for acceleration in the total UNION2.1 sample $13.6 \sigma$ is in accordance with the simulated $\Lambda$CDM universe $12.77 \pm 0.99 \sigma$. By excluding the low-z ($z<0.2$) sub-sample the evidence reachs $17.0 \sigma$. Comparing to the results obtained in Refs. \cite{Seikel:2007pk,Seikel:2008ms} the recent catalogues present even more robust evidence favoring acceleration.


\begin{table}
\centering\caption{(Averaging over SNe Ia) Evidence in the UNION 2.1.  }
{\begin{tabular} {c||c||c}
Redshift & Evidence  in $\sigma$ of C.L. & \# SNe Ia \\
bin & $H_0=73.24 \, (H_0=70.00)$& in the bin \\
\hline \hline
0.0 - 0.2 & 14.1 (4.1) & 230 \\
0.2 - 0.4 & 17.1 (10.0) & 125 \\
0.4 - 0.6 & 14.1 (9.5)& 101  \\
0.6 - 0.8 & 10.6 (7.3) & 51  \\
0.8 - 1.0 & 7.5 (5.2) & 44  \\
1.0 - 1.2 & 6.9 (5.0) & 16  \\
1.2 - 1.41 & 7.5 (5.8) & 13  \\
Total \# of SN & 26.3 (13.6) & 580  \\
0.2 - 1.141 & 26.3 (17.0) & 350  \\
\hline
\end{tabular}} 
\label{TableAveUNION}
\end{table}

\section{The JLA data set}

The Joint Light-Curve analysis (JLA) \cite{Betoule:2014frx} totals 740 SNe Ia including several low-redshift samples $(z<0.1)$, the SDSS-II data $(0.05<z<0.4)$, three years data from SNLS $(0.2<z<1)$ and a few high redshift from the Hubble Space Telescope (HST).

The free parameters fitted in the observed distance modulus for the catalog we use are $\alpha=0.141$, $\beta=3.101$, $M_{B}=-19.05$ and $\delta=-0.070$ which have been fitted with the $\Lambda$CDM cosmology.
 
\begin{figure}
\centering
\includegraphics[width=0.48\textwidth]{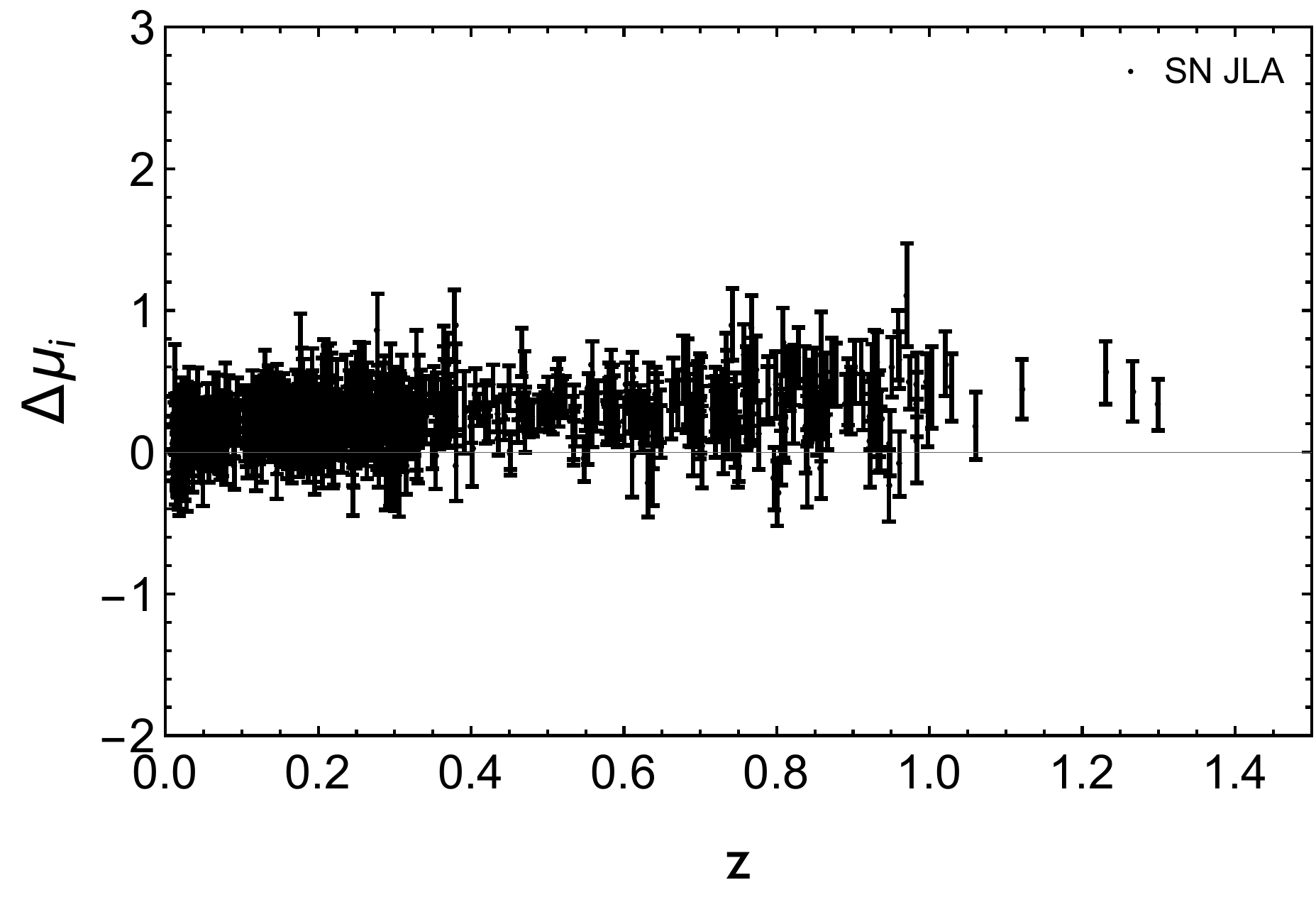}
\caption{Single SNe Ia analysis: $\Delta \mu_i$ for each supernova (adopting $H_0=73.24$ km s$^{-1}$ Mpc$^{-1}$) in the JLA data set.}
\label{FigSingleJLA}
\end{figure}

With such JLA sample we perform the single SNe Ia analysis as one can see the results in Fig. \ref{FigSingleJLA} and Table \ref{TableSingleJLA}. Clearly, this sample presents a smaller dispersion than the UNION2.1 but still with a similar signal-to-noise ratio. 

The averaged analysis per bin in the JLA sample is presented in Table (\ref{TableAveJLA}). In order to study the effects of nearby SNe Ia we both exclude again all SNe Ia at redshifts $z<0.2$ as well as the SNe belonging to the $low-z$ sub-catalog. 

\begin{table}
\centering\caption{(Single SNe Ia Analysis) Number of JLA SNe Ia indicating acceleration or deceleration. The values in parenthesis are calculated using $H_0=70$ km s$^{-1}$ Mpc$^{-1}$.}
{\begin{tabular} {c||c||c}
Dynamics & JLA & JLA  \\
 &  & (without low-z) \\
\hline \hline 
Acceleration-$95 \% C.L.$ & 266 (131) & 252 (129) \\
Acceleration-$99 \% C.L.$ & 115 (43) & 113 (42) \\
Deceleration-$95 \% C.L.$ & 0 (7) & 0 (3)  \\
Deceleration-$99 \% C.L.$ & 0 (0) & 0 (0)  \\
Total number of SN & 740 & 622  \\
\hline
\end{tabular}} 
\label{TableSingleJLA}
\end{table}


\begin{table}
\centering\caption{(Averaging over SNe Ia) Evidence in the JLA.  }
{\begin{tabular} {c||c||c}
Redshift & Evidence  in $\sigma$ of C.L. & \# SNe Ia \\
bin & $H_0=73.24 \, (H_0=70.00)$& in the bin \\
\hline \hline
0.0 - 0.2 & 21.2 (8.8) & 318 \\
0.2 - 0.4 & 21.5 (12.4) & 207 \\
0.4 - 0.6 & 18.3 (12.2) & 70  \\
0.6 - 0.8 & 15.1 (10.3) & 78  \\
0.8 - 1.0 & 11.3 (8.2) & 59  \\
1.0 - 1.299 & 9.3 (7.2) & 8  \\
Total number of SN & 36.7 (20.4) & 740  \\
0.2 - 1.299 & 33.1 (21.2) & 422  \\
Without Low-z & 37.4 (22.1) & 622\\
\hline
\end{tabular}} 
\label{TableAveJLA}
\end{table}



\section{Final discussion}

Rather than using the traditional approach of fitting SNe Ia data with the $\Lambda$CDM model to assess the best-fit values of the cosmological parameters we have studied the late-time cosmic acceleration with a model-independent estimator $\Delta \mu_{obs}$.  

The essence of this estimator is to falsify the {\it null hypothesis} that the universe never expanded in an accelerated way. From our analysis with mock catalogs in section 3.1 we have found however that the estimator actually provides an averaged balance between the accelerated and decelerated periods. Although the models based on the equation of state parameter (\ref{wzt}) can be seen as unrealistic expansions, they serve as counterexamples to show that if data has such untypical distribution the estimator would fail in providing evidence for acceleration in cosmologies that experienced an accelerated epoch. This is in fact related to the fact that $\Delta \mu$ can not be mapped into the deceleration parameter $q(z)$. Therefore, the message here is that one should carefully use this estimator.

Following the spirit of Ref. \cite{Seikel:2007pk,Seikel:2008ms} and assuming that the estimator can be used for a $\Lambda$CDM-like distribution of data as provided by the UNION2.1 and JLA samples, we have also updated the evidence for acceleration obtained from recent catalogs. For the JLA data set we have found robust evidence (see Table \ref{TableAveJLA}) favoring acceleration in a flat FLRW expansion.

It is also evident the strong dependence of the estimator on $H_0$. The larger the $H_0$ value adopted for the estimator, stronger is the evidence favoring acceleration. All reasonable values for $H_0$ lead to positive statistical confidence favoring acceleration.

However, it is worth noting that the UNION2.1 and JLA sample used here with fixed light curve parameters $\alpha,  \beta$, $M_{B}$ and $\delta$ are not actually model-independent since the $\Lambda$CDM model has been adopted in the data fitting. Then, unless the light curve parameters are obtained in a pure astrophysical manner (in the sense their values do not depend on the cosmology adopted) this analysis also can not be regarded as a model-independent one. Attempts to do that by using $H(z)$ data from cosmic chronometers are possible \cite{javier2016}, although systematics regarding the stellar population synthesis model are not negligible \cite{busti2014}. The estimator can be adapted to the $H(z)$ data via the inequality \ref{hzcondition}. We also checked with the 36 data points for $H(z)$ compiled in Ref. \cite{Yu:2017iju} that the evidence favoring acceleration becomes $ 12.1 \, \sigma \,(H_0=68.31), 17.8 \,\sigma \,(H_0=70.00)$ and $23.3 \,\sigma \,(H_0= 73.24)$. Again, though this result clearly shows how the estimator has a strong dependence on $H_0$, there is no doubt about the cosmic acceleration even for lower $H_0$ values.

A more reliable test would verify acceleration independently on the light curve parameters or properly taken into account them. In fact, the light-curve parameters could be even non constant for all SNe. Recent investigations suggest a non trivial dependence of the of the stretch-luminosity parameter $\alpha$ and the color-luminosity parameter $\beta$ on the redshift \cite{Li:2016dqg} or with respect to the host galaxy morphology \cite{Henne:2016mkt}.

A future perspective for our work is the development of a new estimator for assuring cosmic acceleration in a full model-independent way.

\section*{Acknowledgments}

HV and SG thank CNPq and FAPES for partial support. VCB is supported
by FAPESP/CAPES agreement under grant number 2014/21098-1 and FAPESP under grant number 2016/17271-5.

\end{document}